\documentclass[aps,prd,twocolumn,floatfix,superscriptaddress]{revtex4}
\usepackage{amsmath}
\usepackage[dvips]{graphicx}
\usepackage{color} 
\newlength{\colw}
\newcommand{\Ncfg}{N_{\text{cfg}}}

\newcommand{\order}[1]{\mathcal{O}(#1)}

\newcommand{\om}{\omega}

\newcommand{\tmin}{\tau_{\text{min}}}

\newcommand{\ctune}{\cite{Morrin:2006tf}}
\newcommand{\cquench}{\cite{Umeda:2002vr,Asakawa:2003re,Datta:2003ww,Jakovac:2006sf}}

\newcommand{\cspectral}{\cite{Aarts:2007pk}}

\begin{document}

% Definition of title page:
\title{Momentum-dependence of charmonium spectral functions from
  lattice QCD}

\author{Mehmet Bu{\v g}rahan Oktay}
\affiliation{Department of Physics, University of Utah, 115 S 1400 E,
  Salt Lake City, UT 84112-0830 USA.}
\author{Jon-Ivar Skullerud}
\affiliation{Department of Mathematical Physics, NUI Maynooth, County
  Kildare, Ireland.} 

\date{\today}
%\preprint{???}

\begin{abstract}
We compute correlators and spectral functions for $J/\psi$ and
$\eta_c$ mesons at nonzero momentum on anisotropic lattices with
$N_f=2$.  We find no evidence of significant momentum
dependence at the current level of precision.  In the pseudoscalar
channel, the ground state appears to survive up to $T\approx450$MeV or
$2.1T_c$.  In the vector channel, medium modifications may occur at
lower temperatures.
\end{abstract}

\maketitle

%%%%%%%%%%%%%%%%%%%%%% 
\section{Introduction}
\label{sec:intro}
%%%%%%%%%%%%%%%%%%%%%%

The survival and properties of charmonium bound states in the
quark--gluon plasma is a hot topic in the interpretation of
experimental results from relativistic heavy-ion collisions.  A
variety of models have been put forward to explain the observed
pattern of suppression of the $J/\psi$ yield, including sequential
suppression \cite{Karsch:2005nk}, charmonium regeneration through
coalescence of uncorrelated $c\bar{c}$ pairs
\cite{BraunMunzinger:2000px,Thews:2000rj,Thews:2005vj}, and
combinations of these two mechanisms \cite{Grandchamp:2002wp}.

Results from lattice simulations \cquench\ suggest (though far from
decisively \cite{Mocsy:2007yj}) that S-wave ground states ($J/\psi$ and $\eta_c$)
survive up to temperatures close to twice the pseudocritical
temperature $T_c$, while excited states ($\psi'$) and P-waves
($\chi_c$) melt at temperatures close to $T_c$, in rough agreement
with the sequential suppression scenario.  These results are however
not sufficient to rule in favour of one or other scenario.

An additional handle on the problem may be gained by considering
charmonium states which are moving with respect to the thermal
medium.  The two main scenarios (regeneration and sequential
suppression) predict different transverse momentum and rapidity
dependence of the charmonium yield.  This is primarily due to the
momentum dependence of the reaction rates included in these scenarios,
but in order to complete the story it will also be necessary to
investigate any momentum dependence of the baseline survival
probability of charmonium bound states.  This paper provides the first
steps in this direction.

The analytical structure of QCD correlators implies that both spatial
and temporal correlators are governed by the same spectral function.
However, in order to describe spatial correlators and screening
masses, it is necessary to also include the spatial-momentum
dependence of the spectral function.

Finally, computing spectral functions at nonzero momentum may also
facilitate the determination of transport coefficients such as the
heavy quark diffusion constant \cite{Petreczky:2005nh}, as the
zero-frequency transport peak is broadened and shifts to nonzero
frequency.

We will work on a spatially isotropic lattice, which means
that the anisotropy of the plasma created in heavy-ion collisions is
not taken into account.  Our momentum is best interpreted as the
transverse momentum, and we will therefore not be able to say anything
at this point about the rapidity dependence of any observables.

%%%%%%%%%%%%%%%%%%%%%% 
\section{Formulation}
\label{sec:formalism}
%%%%%%%%%%%%%%%%%%%%%%

We have simulated two degenerate flavours of $\order{a}$-improved
Wilson fermions, with $m_\pi/m_\rho=0.54$ (corresponding approximately
to the strange quark mass), and with spatial lattice spacing
$a_s=0.162$fm and anisotropy $\xi=a_s/a_\tau=6$.  The lattice action
is described in more detail in Ref.~\ctune, and the high-temperature
simulation parameters are discussed in Ref.~\cspectral.  Our
parameters here correspond to Run 7 in Ref.~\cspectral.  The same
fermion action is used for both light sea and heavy valence quarks.
The lattices,
corresponding temperatures and number of configurations $\Ncfg$ used
in this study are given in Table~\ref{tab:lattices}.  The main
difference compared to Ref~\cspectral\ is that all our data are now
obtain using the tuned (Run 7) parameters and on a spatial volume of
$12^3$ in lattice units.

\begin{table}[thb]
\begin{tabular}{rcccr}
 $N_s$ & $N_\tau$ & $T$ (MeV) & $T/T_c$ & $\Ncfg$ \\ \hline
 \,8 & 80 & \,92 & 0.42  &  250 \\
 12  & 80 & \,92 & 0.42  &  250 \\
 12  & 32 & 230  & 1.05  & 1000 \\
 12  & 28 & 263  & 1.20  & 1000 \\
 12  & 24 & 306  & 1.40  &  500 \\
 12  & 20 & 368  & 1.68  & 1000 \\
 12  & 18 & 408  & 1.86  & 1000 \\
 12  & 16 & 459  & 2.09  & 1000 \\\hline
\end{tabular}
\caption{Lattice volumes, temperatures and number of configurations
  used in this simulation.  The separation between configurations is
  10 HMC trajectories, except for the $N_\tau=80$ runs where
  configurations were separated by 5 trajectories.  The lattice
  spacings are $a_s\sim 0.162$fm, $a_\tau\simeq 7.35\pm 0.03$GeV.}
\label{tab:lattices}
\end{table}

Spectral functions were computed using the maximum entropy method as
described in \cspectral, with the modified kernel introduced in
Ref.~\cite{Aarts:2007wj}.  The charmonium correlators were computed
using all-to-all propagators with dilution in time, color and varying
levels in space and spin.  The momentum values are given by
$p^2=\left(\frac{2\pi}{a_sN_s}\right)^2n^2$, with $n^2=0,1,2,3,4$,
corresponding to $p=0, 0.66, 0.93, 1.14, 1.32$ GeV.  We consider only the
pseudoscalar ($\eta_c$) and vector ($J/\psi$) channels.  Our
correlators have not been renormalised, so the vertical scale on the
figures presented is arbitrary, but the shape of the curves is not.

%%%%%%%%%%%%%%%%%%%%%% 
\section{Results}
\label{sec:results}
%%%%%%%%%%%%%%%%%%%%%%

\subsection{Zero-temperature spectrum}
\label{sec:T0}

\begin{figure}[t]
\includegraphics*[width=\colw]{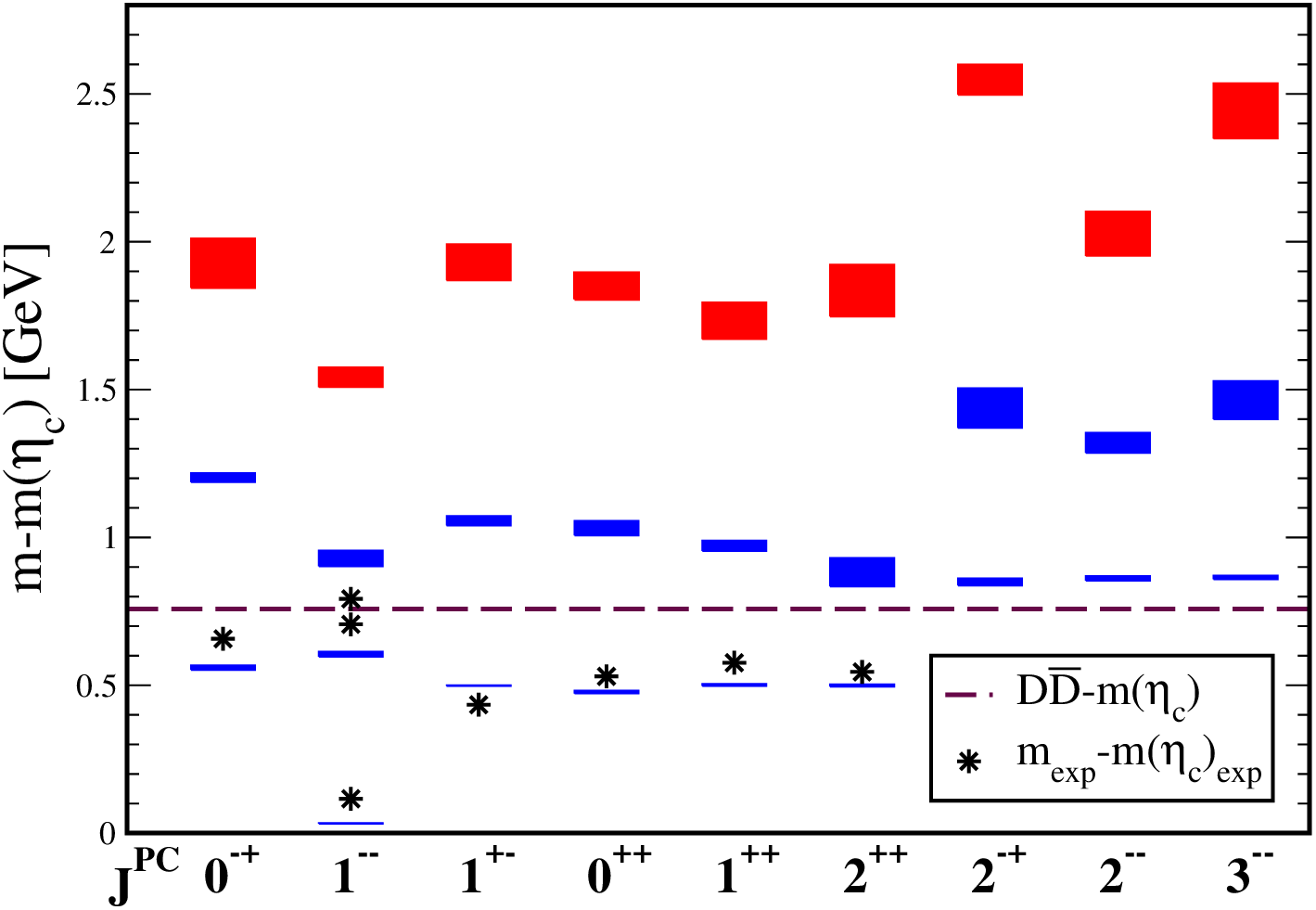}\
\caption{The mass splittings of the low-lying charmonium spectrum at
  zero temperature, with the known experimental values and the
  $D\bar{D}$ threshold also indicated.}
\label{fig:T0spectrum}
\end{figure}

In order to establish a baseline for our high-temperature results, we
first present results for the spectrum and dispersion relation at zero
temperature.  Taking advantage of the all-to-all propagators, we used
the variational method \cite{Luscher:1990ck,Michael:1985ne} with a
large basis of interpolating operators to compute the low-lying
spectrum of S-, P- and D-waves.  This is shown in
Fig.~\ref{fig:T0spectrum}, along with the experimentally known masses.
The operator basis used in the spectrum study is summarized in Table
~\ref{tab:operators}.  Taking the difference between the spin-averaged
1S masses and the $^1\mathrm{P}_1 (h_c)$ mass to set the scale, we
find the inverse lattice temporal lattice spacing to be
$a_\tau^{-1}=7.35(3)$GeV\footnote{This replaces the value 7.23GeV
reported in Ref.~\cspectral, which was obtained using preliminary
values for the 1S and 1P masses.  Our final values for the $\eta_c$
and $J/\psi$ masses are also slightly too high as a result.}.  In
Figure ~\ref{fig:T0spectrum} the highest lying radial excitations in
each channel are colored red to indicate that they are contaminated by
higher excited states.  The difference between our results and
experiment is largely due to the fairly coarse spatial lattice spacing
used.  Including a clover term in the action has been found to improve
the agreement with experiment \cite{Ryan:priv}.
\begin{table}[th]
\begin{tabular}{cccc}
J$^{\mathrm{PC}}$ & ${}^{2S+1}$L$_J$ & State & Operators \\
\hline
$0^{-+}$ & ${}^1S_0$ & $\eta_c,\eta_c^\prime$ & $\gamma_5$,$\gamma_5\sum_is_i$\\
$1^{--}$ & ${}^3S_1$ & $J/\Psi$,$\Psi(2S)$   & $\gamma_j$,$\gamma_j\sum_is_i$\\
%& & & \\  
$1^{+-}$ & ${}^1P_1$ & $h_c$,$h_c^\prime$ & $\gamma_i\gamma_j$,$\gamma_5 p_j$ \\
$0^{++}$ & ${}^3P_0$ & $\chi_{c_0}$,$\chi_{c_0}^\prime$ & 1,
$\vec{\gamma}\cdot\vec{p}$ \\
$1^{++}$ & ${}^3P_1$ & $\chi_{c_1}$,$\chi_{c_1}^\prime$ & $\gamma_5\gamma_i$,
$\vec{\gamma}\times\vec{p}$\\
$2^{++}$ & ${}^3P_2$ & $\chi_{c_2}$,$\chi_{c_2}^\prime$ & 
$\vec{\gamma}\times\vec{p}$, $\gamma_1p_1-\gamma_2p_2$ \\
& & & $2\gamma_3p_3-\gamma_1p_1-\gamma_2p_2$  \\
$2^{-+}$ & ${}^1D_2$ & 1${}^1D_2$ & $\gamma_5(s_1-s_2)$,$\gamma_5(2s_3-s_1-s_2)$
\\
$2^{--}$ & ${}^3D_2$ & 1${}^3D_2$ & $\gamma_j(s_i-s_k)$,
$\gamma_1t_1-\gamma_2t_2$ \\
& & &  $2\gamma_3t_3-\gamma_1t_1-\gamma_2t_2$ \\
$3^{--}$ & ${}^3D_3$ & 1${}^3D_3$ & $\vec{\gamma}\cdot\vec{t}$ \\
\hline
\end{tabular}
\caption{Operator basis used to obtain the $T=0$ charmonium spectrum.
  The definitions of $s_i$,$p_i$ and $t_i$ are given in
  Ref.~\cite{Lacock:1996vy}.}
\label{tab:operators}
\end{table}

\begin{figure}[t]
\includegraphics*[width=\colw]{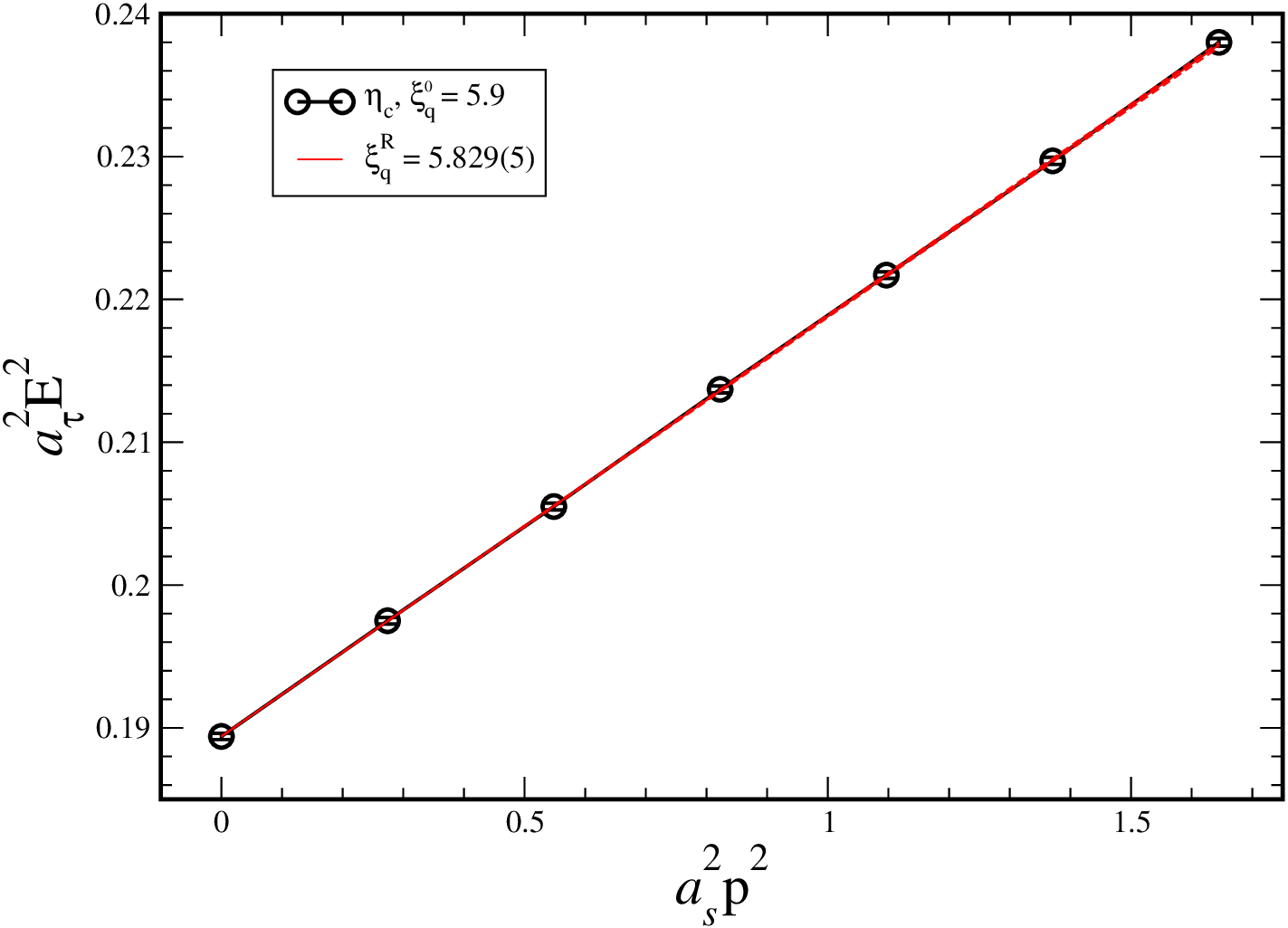}
\caption{The $\eta_c$ dispersion relation.}\label{fig:dispersion}
\end{figure}
Figure~\ref{fig:dispersion} shows the ground-state pseudoscalar
($\eta_c$) dispersion relation.  The bare anisotropy for the valence
charm quarks is $\xi_0^c=5.90$, which on the $8^3\times80$ lattice had
given a renormalised anisotropy of 6.  However, on the larger
($12^3\times80$) lattice we find that the dispersion relation gives us
a renormalised anisotropy of 5.829$\pm$0.05, giving rise to a 3\%
systematic error in our final results.

\subsection{High-temperature spectral functions}
\label{sec:highT}

In order to facilitate comparisons between spectral functions at
different temperatures and momenta, we have determined the spectral
functions $\rho(\om,\vec{p};T)$ at all $T$ and $\vec{p}$ using the
{\em same} default model, $m(\om)=m_0\om^2$ with $a_\tau m_0=6$ and
$a_\tau m_0=4$ for the pseudoscalar and vector channel respectively.  These
values are close to the `best' values determined from a one-parameter
fit of the correlator for each temperature and momentum considered.
We have repeated the analysis for a range of different default models
to study the robustness of our results.

As in Ref.~\cspectral, the first two timeslices were discarded in the
analysis since these will be dominated by lattice artefacts.  The effect
of varying the time range has also been investigated.

\begin{figure}[t]
\includegraphics*[width=\colw]{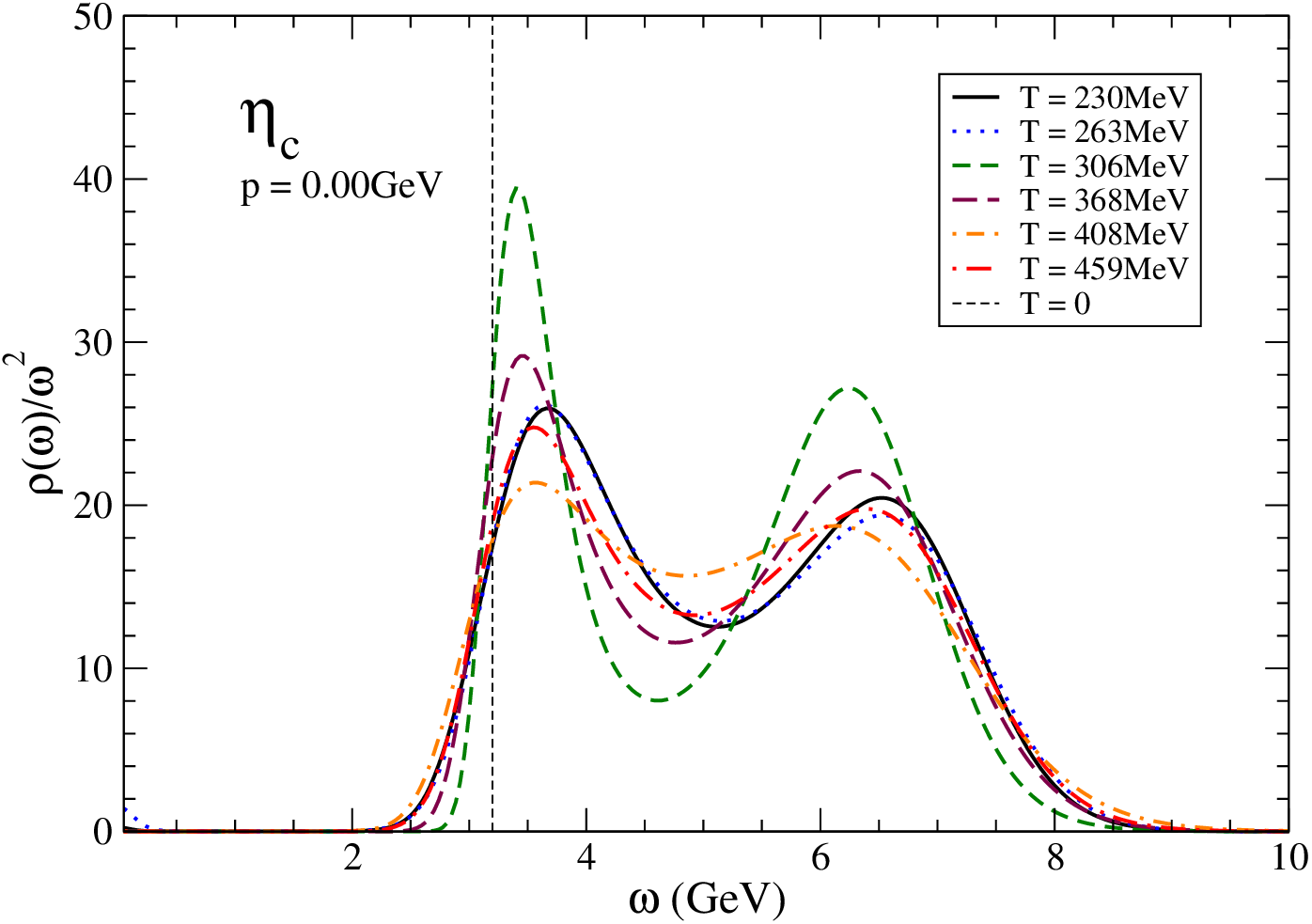}\\
\includegraphics*[width=\colw]{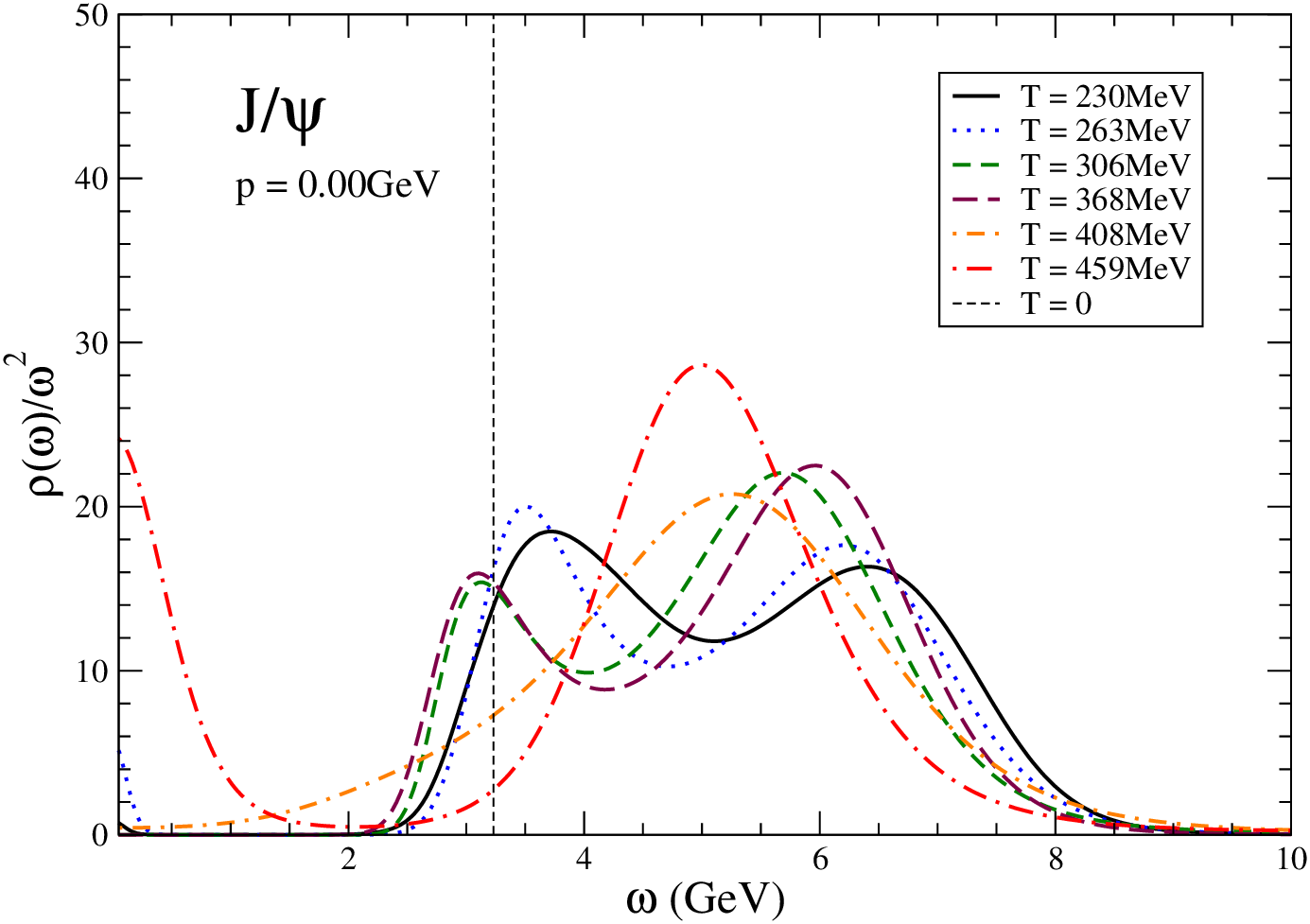}
\caption{The pseudoscalar (top) and vector(bottom) spectral function
  at zero momentum for various temperatures.}\label{fig:p0}
\end{figure}
Figure~\ref{fig:p0} shows the spectral function in both channels at
zero momentum for the various temperatures.  These results may be
compared to Figs. 5 and 6 in Ref.~\cspectral.  In the pseudoscalar
channel, our results suggest that the ground state ($\eta_c$) survives
up to the highest temperatures accessible in our simulations, while
the $J/\psi$ (vector state) appears to dissolve for $T\gtrsim370$MeV
or $1.7T_c$.  It must be noted, however, that the uncertainty in the
MEM procedure increases as the temperature increases, and it is
therefore not possible at present to say for certain whether the
pattern observed at the highest temperatures is genuine or merely
reflects the inability of the MEM algorithm to determine the spectral
function given the available data.  This remains the case for all
momenta and in all channels.

\begin{figure}[t]
\includegraphics*[width=\colw]{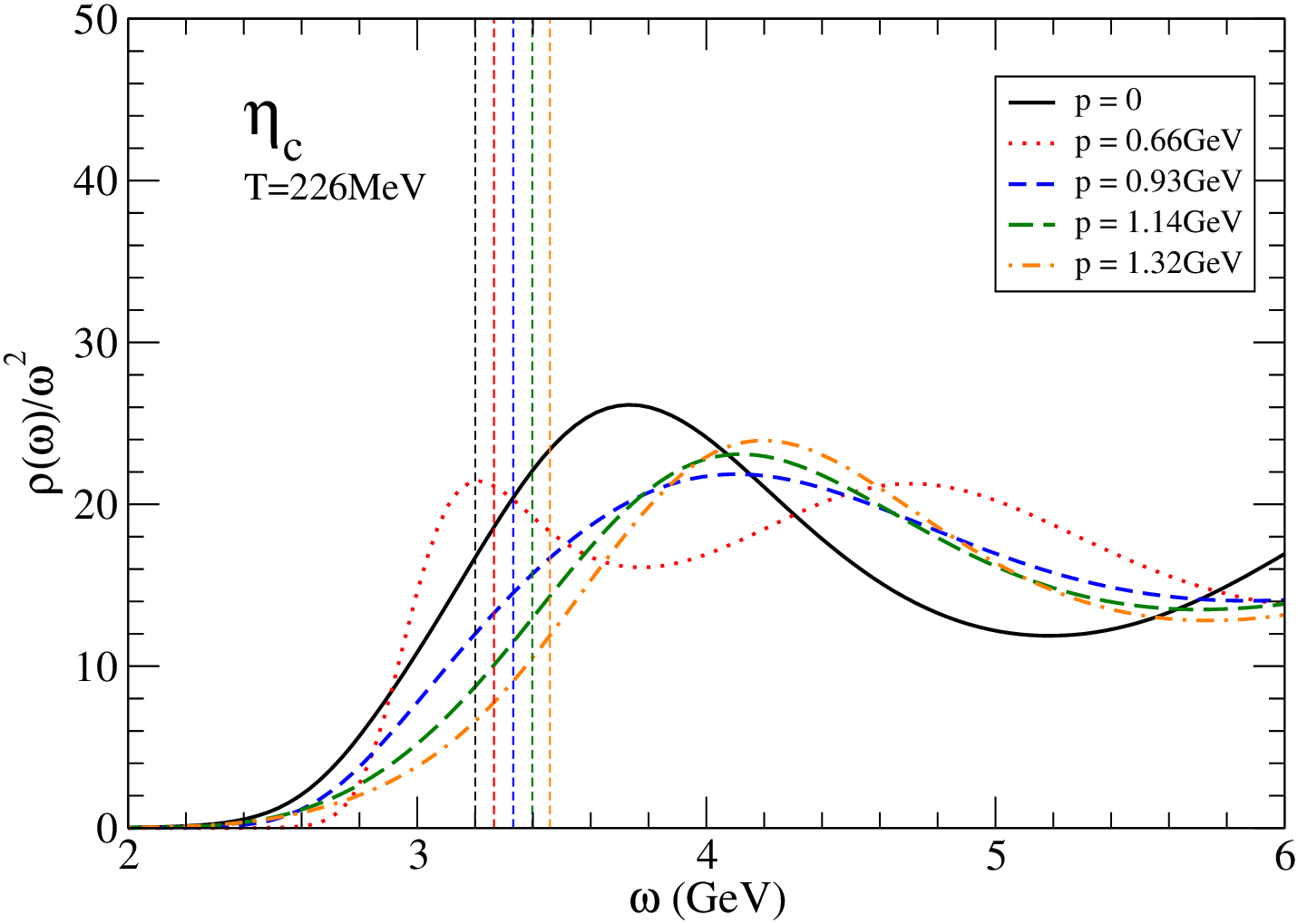}\\
\includegraphics*[width=\colw]{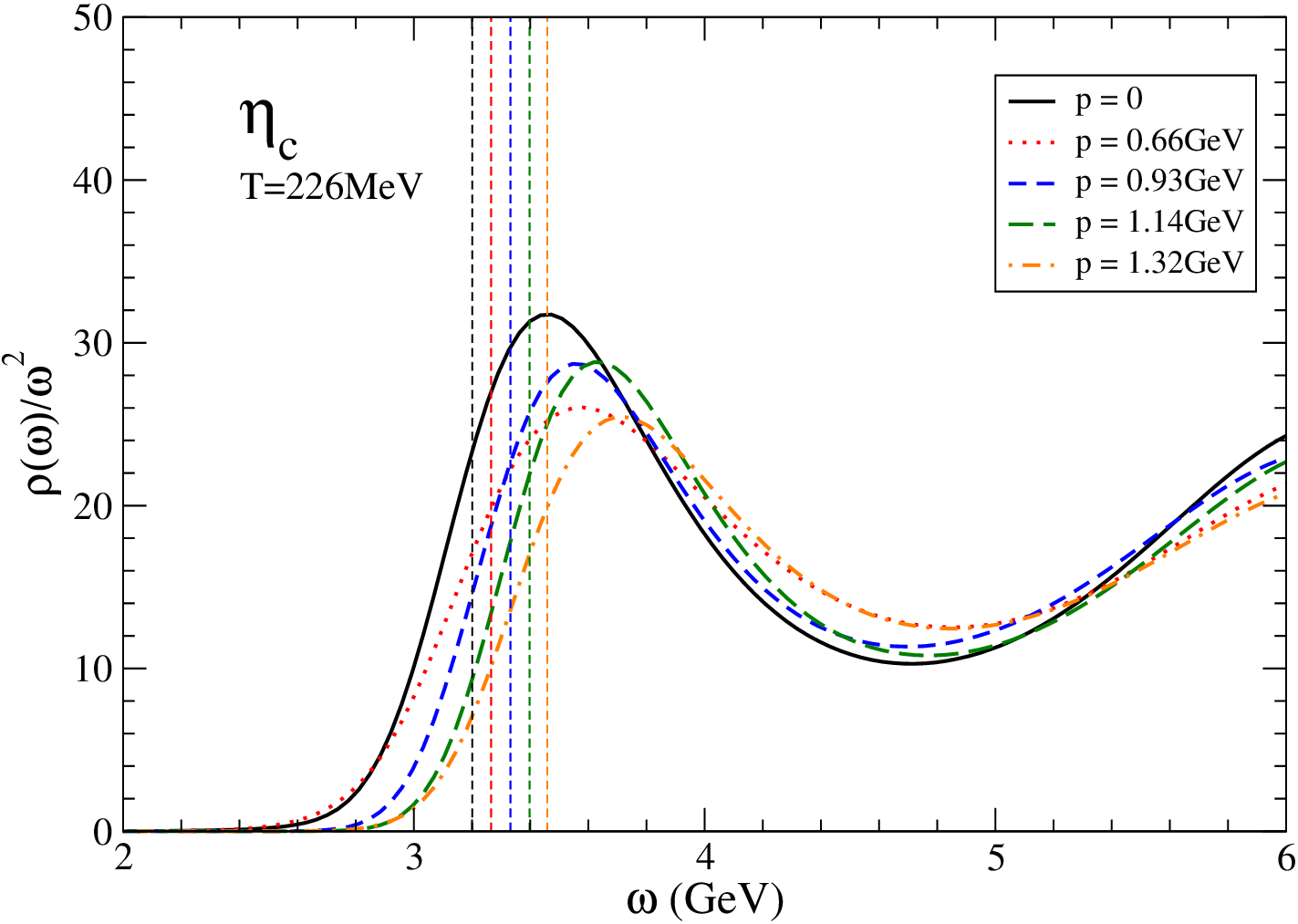}
\caption{The pseudoscalar spectral function at nonzero momentum for
  $T=230$ MeV, using timeslices 2--16 (top) and 3--16 (bottom).  The
  zero-temperature energy levels are also shown for
  comparison.}\label{PS-p-lowT}
\end{figure}
Figure~\ref{PS-p-lowT} shows the pseudoscalar spectral function for
different momenta at $T=230$ MeV.  Using $\tmin=3$ (lower panel), we see
a clear structure of peaks ordered by momentum, with the separation
between the peaks corresponding reasonably well to the
zero-temperature energy levels.  The peak positions, however, appear
to be shifted compared to the zero-temperature energies.  This is most
likely due to the maximum entropy method not being able to resolve the
full detail of the spectral function for the available data.  This is
supported by the upper panel of Fig.~\ref{PS-p-lowT}, showing the
spectral functions determined using $\tmin=2$.  Here the peak position
for the lowest non-zero momentum corresponds precisely to the
zero-temperature energy, while for the other momenta the peaks are
smeared out and shifted to higher energies.

\begin{figure}[t]
\includegraphics*[width=\colw]{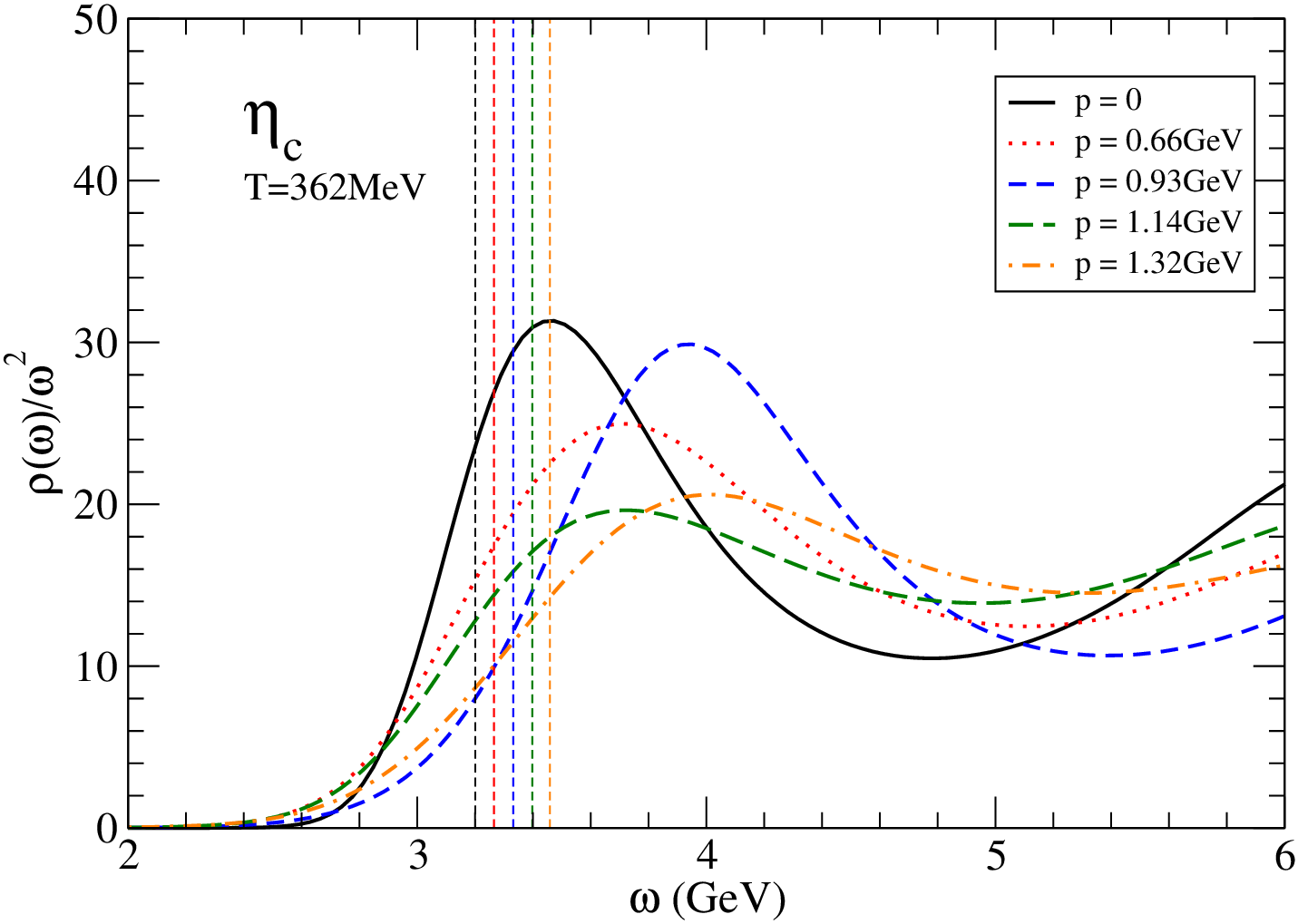}\\
\includegraphics*[width=\colw]{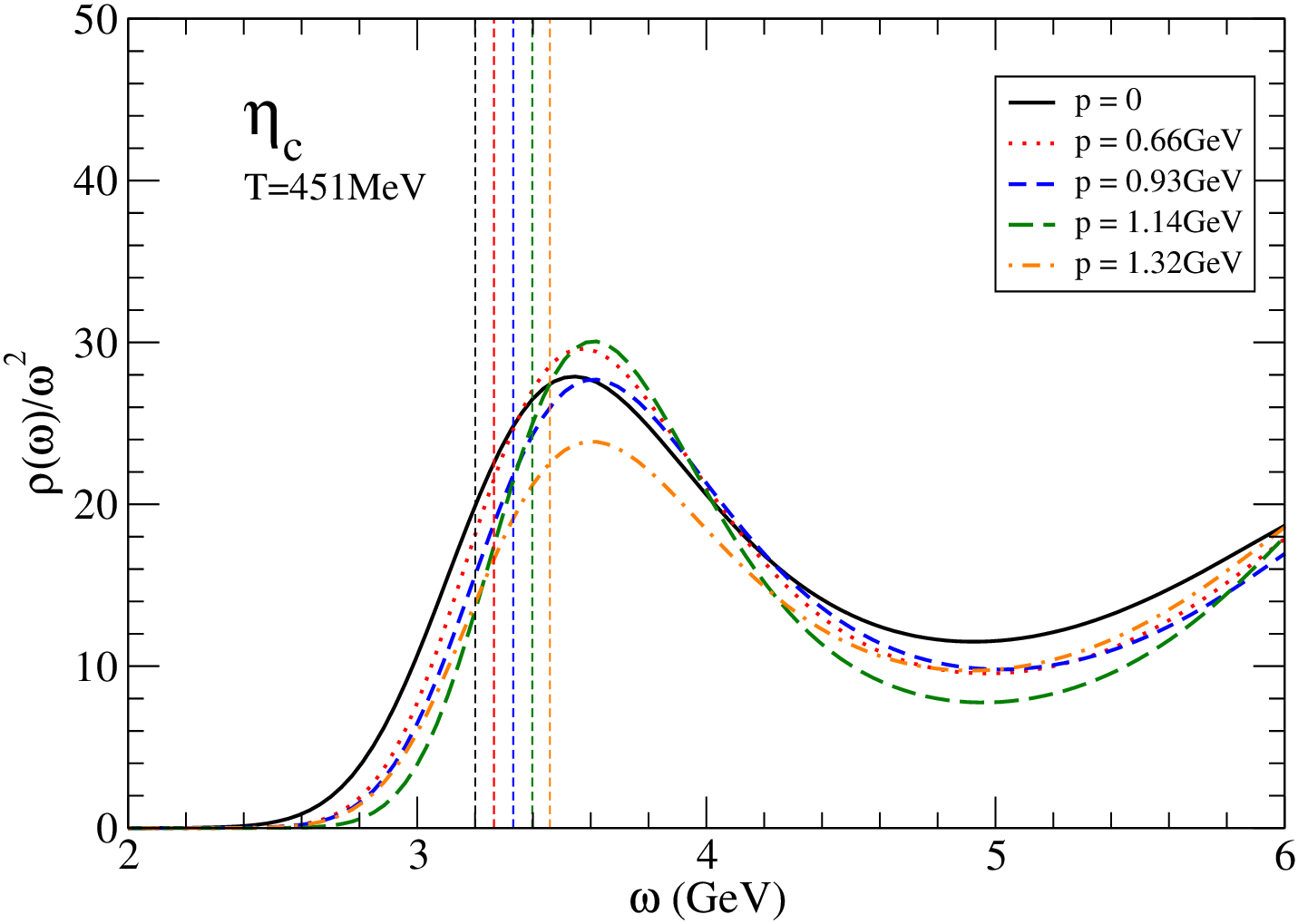}
\caption{The pseudoscalar spectral function at nonzero momentum for
  $T=368$ MeV (top) and $T=459$ MeV (bottom).  The zero-temperature
  energy levels are also shown for comparison.}\label{PS-p-highT}
\end{figure}
Figure~\ref{PS-p-highT} shows the pseudoscalar spectral function for
different momenta at two higher temperatures.  There is still some
evidence of a surviving ground state peak at all momenta, but
uncertainties in the MEM reconstruction means that we are no longer
able to resolve the ordering of the states given our current
precision.

At nonzero momentum, the vector meson correlator is decomposed into 
transverse and longitudinal polarisations,
\begin{equation}
V_{ij}(\tau,\vec{p})
 = \big(\delta_{ij}-\frac{p_ip_j}{p^2}\big)V_T(\tau,\vec{p})
  + \frac{p_ip_j}{p^2}V_L(\tau,\vec{p})\,.
\end{equation}
The transversely and longitudinally polarised $J/\psi$ may in
principle behave differently in the medium.  We have therefore
analysed the two separately.
\begin{figure*}[htb]
\includegraphics*[width=\colw]{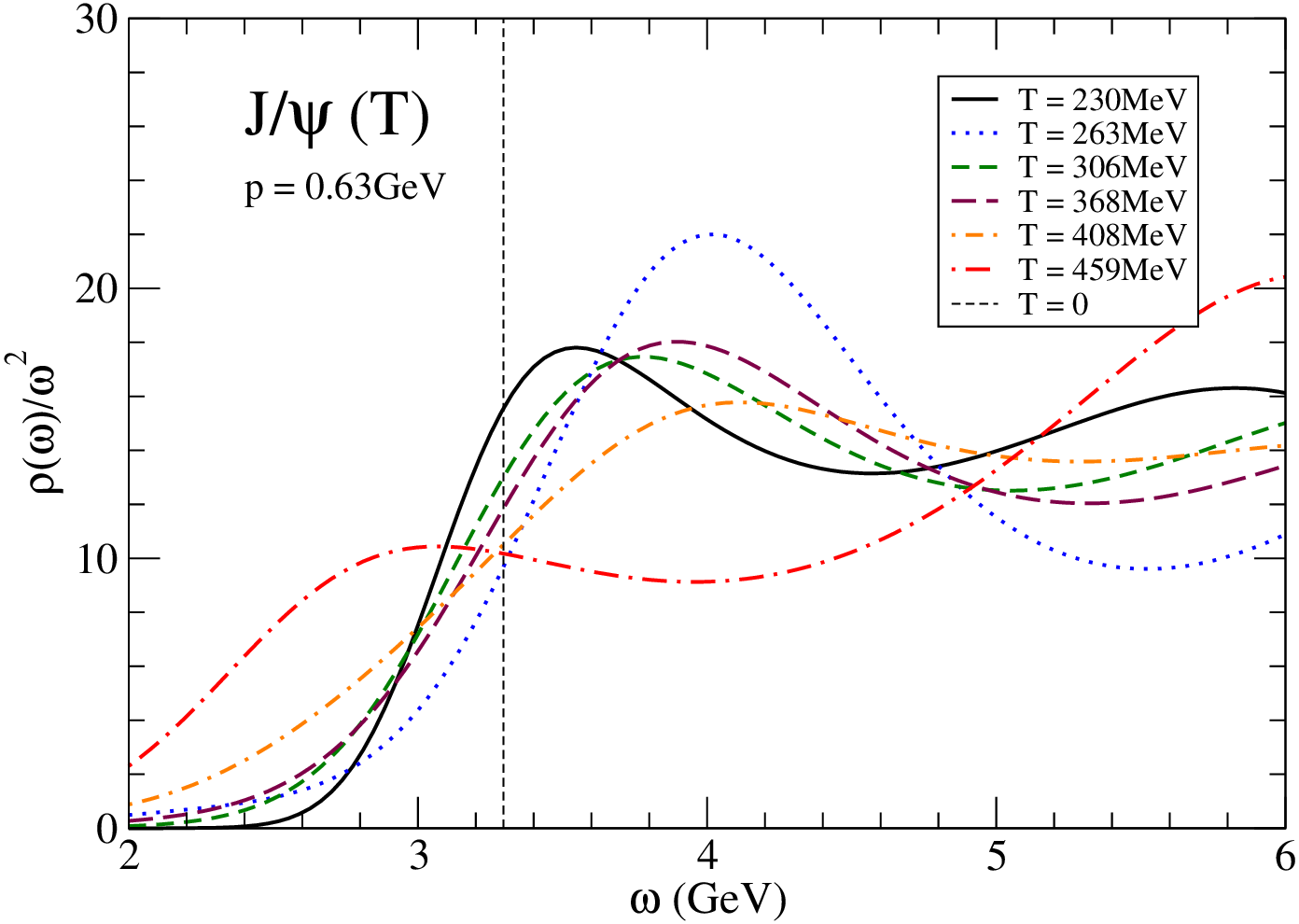}
\includegraphics*[width=\colw]{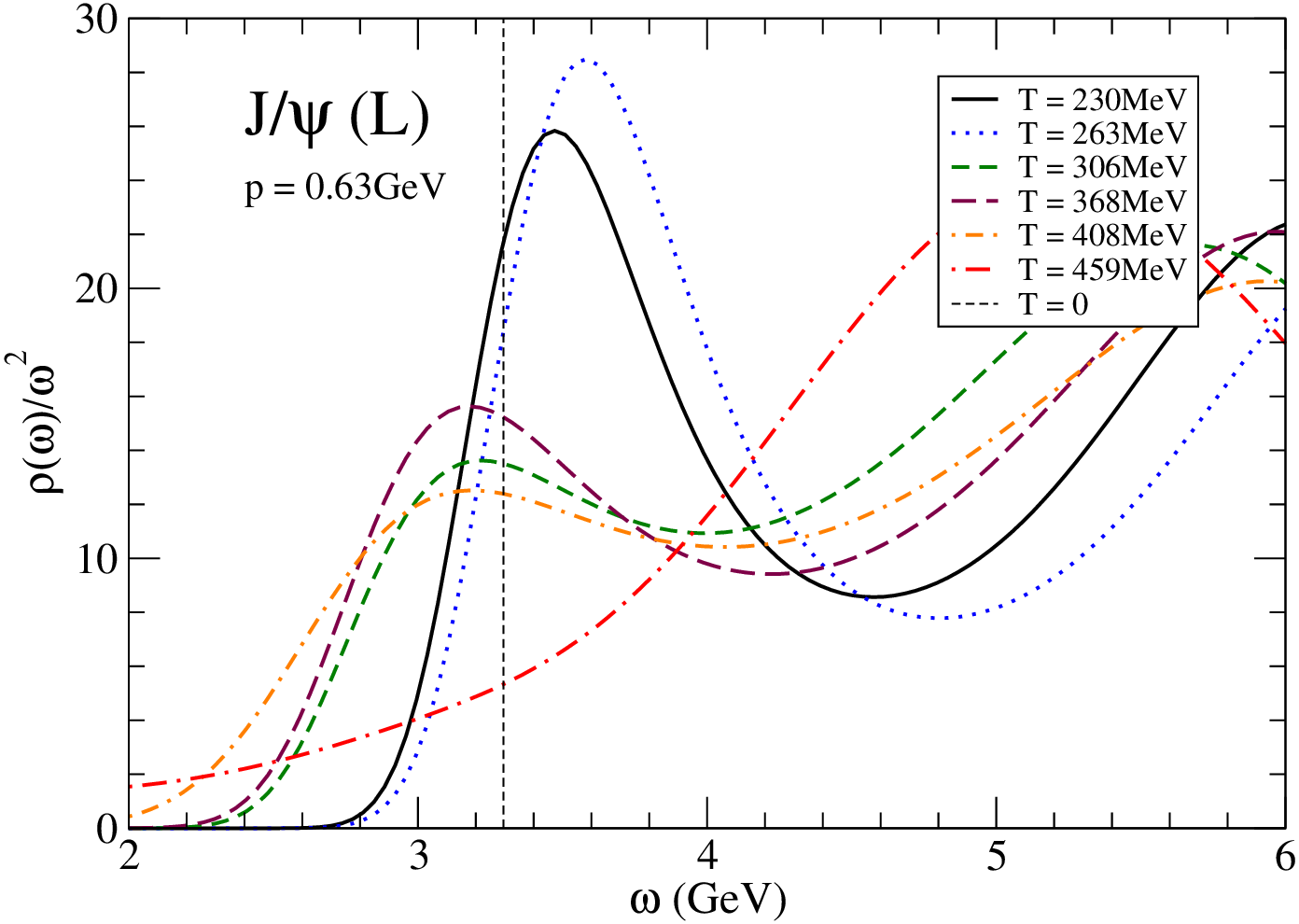}
\caption{The transverse (left) and longitudinal (right) vector meson
  spectral function at $p=0.66$GeV for various temperatures.}\label{V-p1}
\end{figure*}
\begin{figure*}[htb]
\includegraphics*[width=\colw]{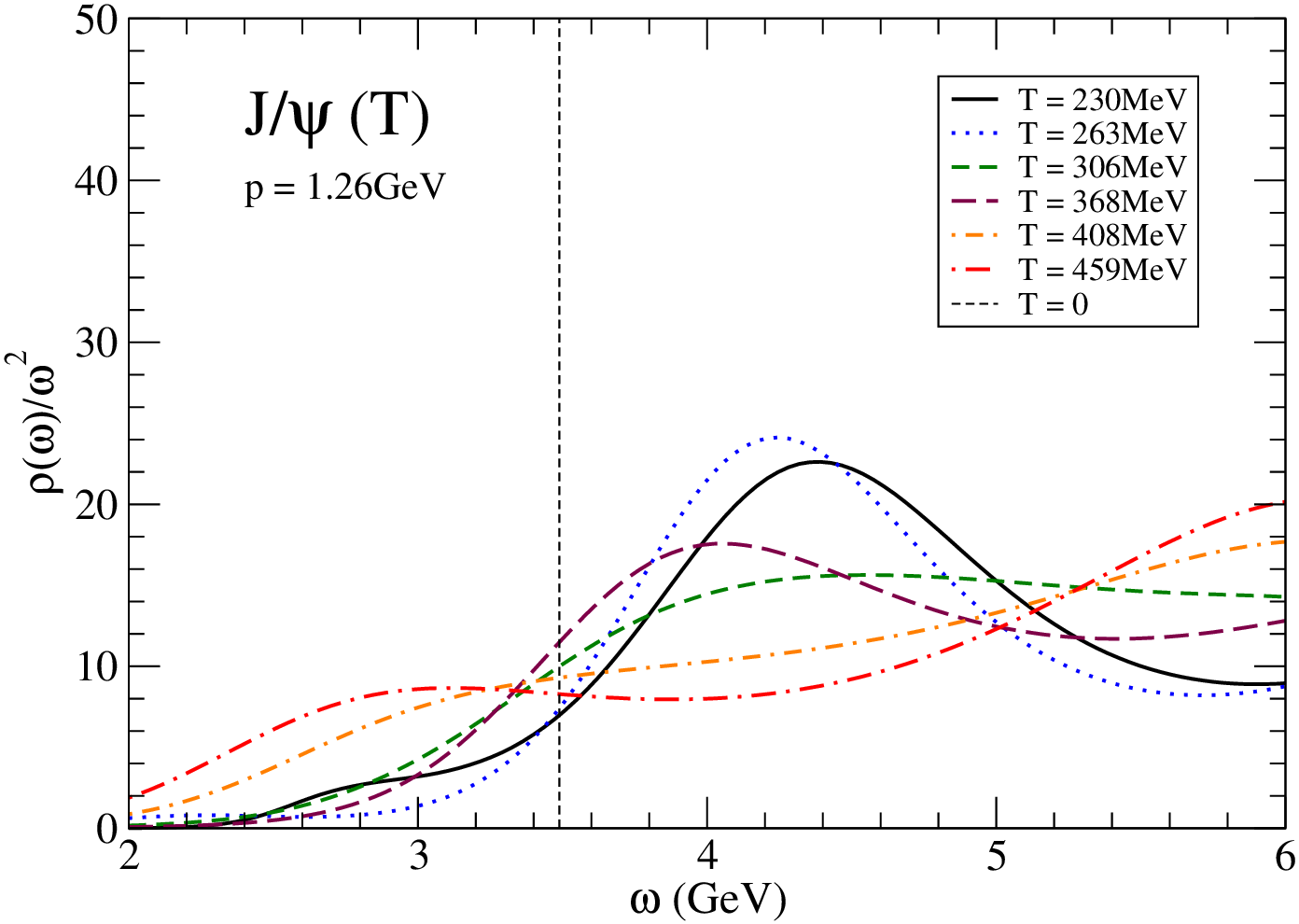}
\includegraphics*[width=\colw]{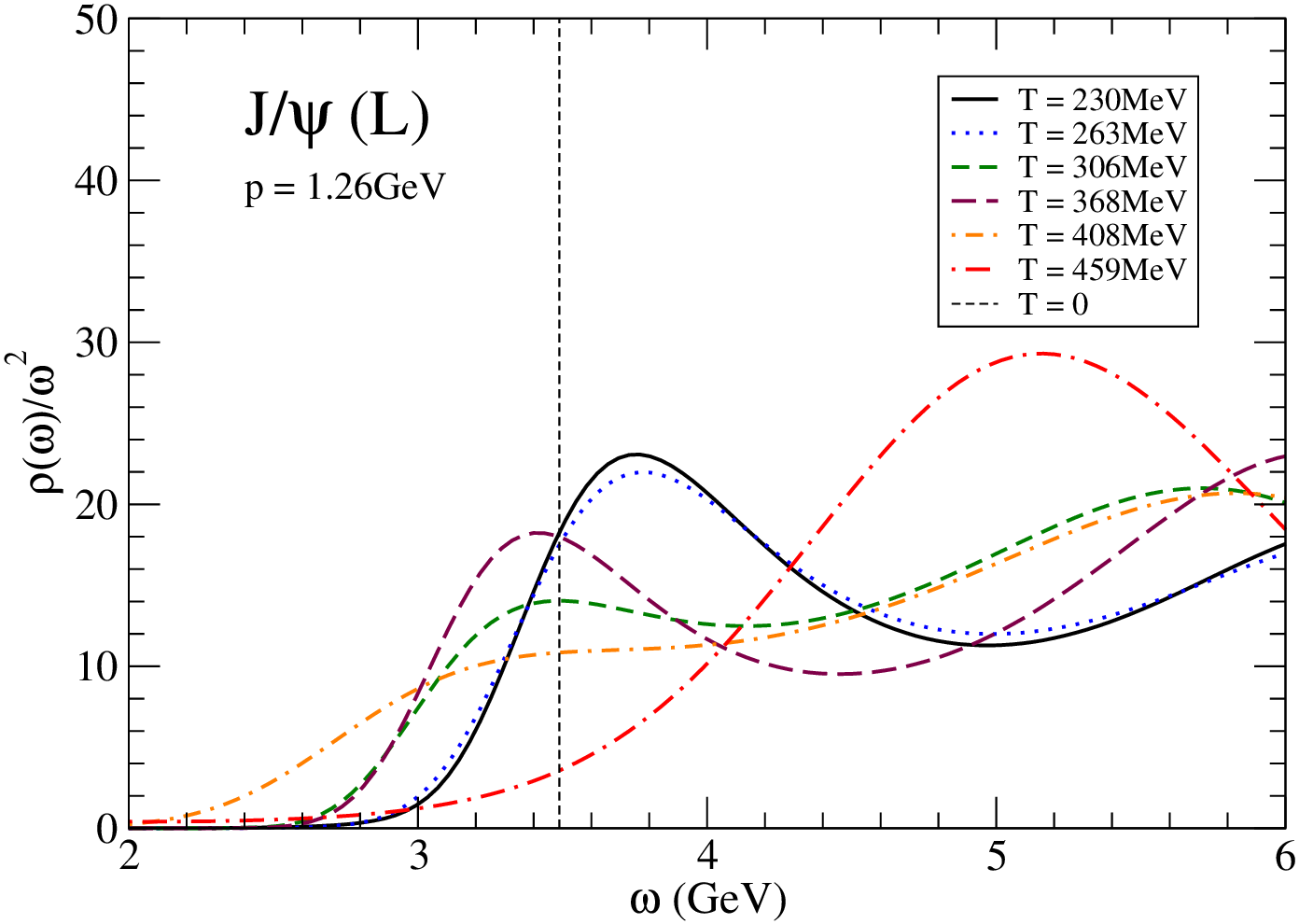}
\caption{The transverse (left) and longitudinal (right) vector meson
  spectral function at $p=1.32$GeV for various temperatures.}\label{V-p4}
\end{figure*}
Figures~\ref{V-p1} and \ref{V-p4} show the transverse and longitudinal
spectral functions for our largest and smallest momentum,
respectively. The longitudinal spectral function for the lowest
momentum ($p=0.66$GeV) shows clear evidence of a surviving peak
corresponding to the $J/\psi$ ground state at the two lowest
temperatures, but there are indications of medium modifications
already at $T=300$MeV.  The transverse spectral function, on the other
hand, appears to be subject to medium modifications (indicated by the
softer ground state peak) already at the lowest temperature, just
above $T_c$.

It is worth emphasising that the transverse spectral
function at $T=230$MeV, $p=0.66$GeV displays little or no sensitivity
to the choice of model function, so we consider our result in this
case to be robust.
At higher momenta and temperatures, the MEM reconstruction of
spectral functions is no longer stable with regard to variations in
the model function and time range used.  Our results for both
transverse and longitudinal $J/\psi$ at $p=1.32$GeV should therefore
be taken as no more than indicative, even at the lowest temperature.
With this proviso, the indications are that the longitudinal $J/\psi$
survives at least up to 265MeV ($1.2T_c$) and is most likely melted by
$T=400$MeV, while the transversely polarised $J/\psi$ may experience
medium modifications already close to $T_c$.

%%%%%%%%%%%%%%%%%%%%%% 
\section{Discussion}
\label{sec:dicuss}

We have presented first results for charmonium spectral functions at
nonzero momentum.  Our results suggest that the 1S pseudoscalar meson
($\eta_c$) survives up to temperatures close to twice the
pseudocritical temperature of QCD, for all momenta.  No substantial
momentum dependence was found.

In the vector channel, there appears to be a distinction between the
transverse and longitudinal channels, with the longitudinally
polarised $J/\psi$ experiencing smaller medium modifications.  Again,
no substantial momentum dependence was found, although the
reconstruction of the spectral function became progressively more
uncertain with increasing momentum.  Therefore, some additional momentum
dependence can not be ruled out.

Whether the difference between transverse and longitudinal
polarisations is real or merely a reflection of the uncertainty in the
MEM procedure given the data used in this study, still needs to be
determined.

We are in the process of generating configurations with
smaller lattice spacing.  This will provide greater temporal
resolution, leading to a more reliable determination of spectral
functions from imaginary-time correlators.  The finer lattice is also
expected to bring the zero-temperature spectrum in closer agreement
with experiment, and allow a clearer separation between physical
features and lattice artefacts in our spectral functions.
It will also allow us to access higher temperatures, where all model
studies up to now have found the charmonium ground states to be
dissolved.

We are also computing correlators of the conserved vector current,
which will provide the correct nonperturbative renormalisation of the
vector operator used in this study, and permit a quantitative
determination of the charm quark diffusion rate and the charmonium
contribution to the dilepton rate.

%%%%%%%%%%%%%%%%%%%%%%

\begin{acknowledgments}
This work was supported by SFI grant RFP-08-PHY1462, U. S. Department
of energy under grant number DE-FC06-ER41446 and by the U. S. National
Science Foundation under grant numbers PHY05-55243 and PHY09-03571.
We are grateful to the Trinity Centre
for High-Performance Computing for their support.
We have benefited greatly from numerous discussions with Sin\'ead Ryan
and Mike Peardon.

\end{acknowledgments}

%\bibliography{trinlat_bib/trinlat,spectral}

\end{document}